\documentstyle[12pt,a4,epsfig]{article}

\newcommand{\pT}{p_{\perp}}
\newcommand{\kT}{k_{\perp}}

\newcommand{\B}{{\mathrm B}}
\renewcommand{\b}{{\mathrm b}}
\renewcommand{\c}{{\mathrm c}}
\renewcommand{\d}{{\mathrm d}}
\newcommand{\e}{{\mathrm e}}
\newcommand{\g}{{\mathrm g}}
\newcommand{\p}{{\mathrm p}}
\newcommand{\q}{{\mathrm q}}
\renewcommand{\B}{{\mathrm B}}
\newcommand{\Z}{{\mathrm Z}}
\newcommand{\bbar}{\overline{\mathrm b}}
\newcommand{\cbar}{\overline{\mathrm c}}

\newcommand{\pbar}{\overline{\mathrm p}}
\newcommand{\mb}{m_{\mathrm b}}
\newcommand{\mc}{m_{\mathrm c}}
\newcommand{\JP}{{\mathrm J}/\psi}
\newcommand{\LQCD}{\Lambda_{\mathrm QCD}}
\newcommand{\as}{\alpha_{\mathrm s}}
\newcommand{\PRL}[1]{{\it Phys.\ Rev.\ Lett.\ }{\bf #1}}
\newcommand{\PRD}[1]{{\it Phys.\ Rev.\ }{\bf D#1}}
\newcommand{\NPB}[1]{{\it Nucl.\ Phys.\ }{\bf B#1}}
\newcommand{\NPA}[1]{{\it Nucl.\ Phys.\ }{\bf A#1}}
\newcommand{\PLB}[1]{{\it Phys.\ Lett.\ }{\bf B#1}}
\newcommand{\ZPC}[1]{{\it Z.\ Phys.\ }{\bf C#1}}
\newcommand{\hep}[1]{{hep-ph/#1}}

\newlength{\abstwidth}
\begin{document}

\sloppy

\renewcommand{\arraystretch}{1.5}

\pagestyle{empty}

\begin{flushright}
CERN--TH/98--132\\ 
hep-ph/9804349
\end{flushright}
 
\vspace{\fill}

\begin{center}
{\Large{\bf 
Testing factorization of charmonium production}}\\[10mm]
{\Large Gerhard A. Schuler$^a$} \\[3mm]
{\it Theory Division, CERN,} \\[1mm]
{\it CH-1211 Geneva 23, Switzerland}\\[1mm]
{ E-mail: Gerhard.Schuler@cern.ch}
\end{center}
 
\vspace{\fill}
\begin{center}
{\bf Abstract}\\[2ex]
\begin{minipage}{\abstwidth}
Within the NRQCD factorization approach the production of heavy quarkonia 
can be calculated in perturbative QCD in terms of a few 
production-process-independent NRQCD matrix elements. 
We test the universality of these long-distance matrix elements 
by comparing their values as determined from 
charmonium production at $\p\pbar$, $\e\p$, and $\e^+\e^-$ 
colliders, and in $\Z^0$, $\Upsilon$, and $\B$ decays.
\end{minipage}
\end{center}

\vspace{\fill}
\noindent
\rule{60mm}{0.4mm}

%
\vspace{3mm} \noindent
${}^a$ Heisenberg Fellow.

\vspace{10mm}\noindent
CERN--TH/98--132 \\
April 1998

\clearpage
\pagestyle{plain}
\setcounter{page}{1} 


\section{Introduction}
The production of heavy quarkonia in high-energy reactions can
be calculated in perturbative QCD in terms of a few 
non-perturbative parameters that are independent of the production process. 
More precisely, a factorization theorem holds, which expresses 
the cross section to produce a quarkonium $H$ as a series of terms
\begin{equation}
  \sigma\left( H \right) = 
 \sum_{n}\, 
  \sigma\left( Q\bar{Q}[n] \right)\, \langle {\cal O}^H[n] \rangle
\ ,
\label{factcross}
\end{equation}
where $\sigma( Q\bar{Q}[n] )$ is the cross section to produce 
a quark--antiquark pair with small relative momentum 
at distances $1/m$ or smaller ($m$ is the heavy-quark mass). 
The quantity $ \langle {\cal O}^H[n] \rangle$ gives the probability 
for the heavy-quark pair of state $n$ to bind into the meson $H$. 
On the one hand, the binding occurs at distances much larger than
$1/m$, ensuring the existence of the factorized form (\ref{factcross}). 
On the other hand, the binding energy for neither charmonia nor bottomonia 
is large compared with the dynamical low-energy QCD scale $\LQCD$. 
Hence the long-distance matrix elements (MEs) 
$ \langle {\cal O}^H[n] \rangle$ 
cannot be computed perturbatively. However, there
exists a power-counting scheme 
that provides a hierarchy of the MEs so that to any desired accuracy
only a given number of terms contribute in (\ref{factcross}).

The factorization (\ref{factcross}) has been derived perturbatively 
using the so-called non-relativistic QCD (NRQCD), an effective field theory
appropriate for mesons containing a heavy-quark pair \cite{BBL95}.
It therefore holds 
to leading order in $\LQCD/m$. Higher-twist corrections occur
both in the production of the heavy-quark pairs and in their subsequent 
fragmentation into the meson. The importance of the NRQCD MEs 
$ \langle {\cal O}^H[n] \rangle$ is governed by their scaling
with $v = |\vec{v}|$, the relative velocity between the heavy quark $Q$ 
and the heavy antiquark $\bar{Q}$ within the bound state. 
Equation~\ref{factcross} is hence a double expansion in $v^2$
and $\as$ ($ = \as(\mu)$ with $\mu \sim m$). The occurrence of $v^2$
as an expansion parameter is in line with the success of
non-relativistic potential models in describing quarkonium spectroscopy 
and distinguishes quarkonia from heavy--light systems where 
$\LQCD/m$ is the expansion parameter. 

In the non-relativistic limit the quarkonium is a pure $Q\bar{Q}$ state 
with quantum numbers that match those of the meson, 
for example a $\c\cbar_1(^3S_1)$ state in the case of the $\JP$ 
($J^{PC}=1^{--}$) 
and a $\c\cbar_1(^3P_J)$ state in the case of the $J^{++}$ $\chi_{\c J}$ 
mesons. The phenomenologically most important consequences of the
NRQCD factorization approach (FA) are two-fold. First, 
two MEs contribute to $\chi_{\c J}$ production at leading order in $v^2$, 
namely 
$ \langle {\cal O}_1^{\chi_{\c J}}(^3P_J) \rangle$ and
$ \langle {\cal O}_8^{\chi_{\c J}}(^3S_1) \rangle$. That is to say, 
besides the contribution from the leading Fock state, the colour-singlet
$\c\cbar(^3P_J)$ state, there is, at leading order in $v^2$, 
a contribution where the heavy-quark pair is produced in a 
colour-octet state.

The second, perhaps more dramatic, consequence concerns the $1^{--}$ 
mesons ($\JP$, $\psi'$, $\Upsilon(nS)$): although there is just 
one contribution corresponding to the short-distance production of the
leading $Q\bar{Q}_1(^3S_1)$ Fock state for $v^2 \rightarrow 0$, relativistic
corrections are sizeable, even partly dominating the production cross section, 
at least in the charm system: $v^2 \sim 0.3$ is not small enough 
for gluon radiation from an almost point-like $Q\bar{Q}$ pair 
produced in the hard interaction (governed by $\as$)
to dominate gluon radiation from the pair at later times, when it
has already expanded to the quarkonium size $\sim 1/(mv)$
(governed by $v^2$); compare, for example, 
(\ref{ppsingtwo}) and (\ref{pptwo}) below.
Therefore contributions to the production of $1^{--}$ mesons 
are important where 
the heavy-quark pair is produced at short distances in a colour-octet state.
Colour-octet MEs first appear at relative order $v^4$. Different 
quarkonium-production reactions differ in the relative weights in which
these MEs enter the $1^{--}$ cross sections, i.e.\ by their short-distance 
parts $\sigma( Q\bar{Q}[n] )$ with 
$n={}^3S_1^{(1)}$, ${}^3S_1^{(8)}$,  ${}^1S_0^{(8)}$,  and ${}^3P_J^{(8)}$ 
up to and including $O(v^4)$.

\section{Charmonium production at the Tevatron}
\label{sec:tevatron}
In $\p\pbar$ collisions at high energies, charmonium particles come
from the decay of $\b$-flavoured hadrons and prompt production,
the latter consisting of direct production and feed-down from 
higher charmonium states (except for $\psi'$, the highest-lying 
charmonium state). 
Short-distance processes that produce direct $\psi(nS)$'s in the 
leading colour-singlet state are
\begin{eqnarray}
  \g + \g & \rightarrow & \c\cbar_1(^3S_1) + \g 
\ : \qquad \as^3\, \frac{1}{\pT^8}
\ ,
\label{ppsingone}
\\
  \g + \g & \rightarrow & \g + \g^\star\ , 
  \quad \g^\star \rightarrow \c\cbar_1(^3S_1) + \g + \g
\ : \qquad \as^5\, \frac{1}{\pT^4}
\ ,
\label{ppsingtwo}
\end{eqnarray}
where the power of $\as$ and the $\pT$ fall-off (at $90^\circ$ in
the partonic c.m.s.) have been indicated. At large $\pT$, 
(\ref{ppsingtwo}) dominates and can be calculated by folding 
the $\g\g \rightarrow \g X$ subprocess with the $\g \rightarrow \psi X$ 
fragmentation function \cite{BY93}. 
With values of the colour-singlet MEs as obtained
from potential model calculations, (\ref{ppsingone}) and (\ref{ppsingtwo})
underestimate the observed yield of direct $\JP$ and $\psi'$ 
mesons by a factor $\sim 30$. 
\begin{figure}[tbp]
\begin{center}
\epsfig{file=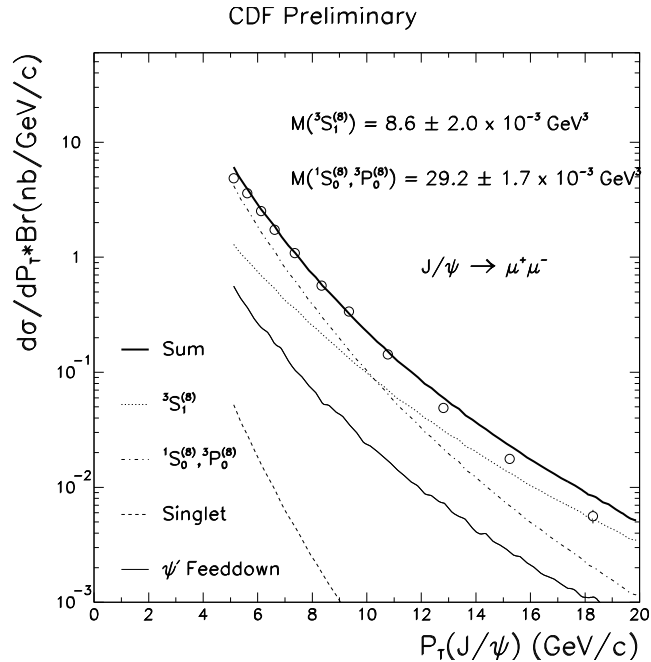,%
width=0.60\textwidth}
\end{center}
\caption{Fits to direct $\JP$ production at the Tevatron 
\cite{ref:tevatron}. 
\label{fig:Tevatron}}
\end{figure}
The data can, however, be accounted for (Fig.~\ref{fig:Tevatron})
if the short-distance production
of colour-octet $Q\bar{Q}$ states is included \cite{pprefs,Cacciari}
\begin{eqnarray}
  \g + \g & \rightarrow & \c\cbar_8[n] + \g \qquad 
     n = {}^1S_0\ , {}^3P_J
\ : \qquad \as^3\, v^4\, \frac{4m^2}{\pT^6}
\ , 
\label{ppone}
\\
  \g + \g & \rightarrow & \g + \g^\star \ , \qquad 
   \g^\star \rightarrow \c\cbar_8({}^3S_1)
\ : \qquad \as^3\, v^4\, \frac{1}{\pT^4}
\ .
\label{pptwo}
\end{eqnarray}
The $\pT$ shapes of these two mechanisms are sufficiently different
over the measured range to isolate the $^3S_1$ colour-octet ME, 
but the remaining two MEs can be extracted only in the linear combination
\begin{equation}
  M_a^H =   \left\langle {\cal O}_8^{H}(^1S_0) \right\rangle 
 + \frac{a}{\mc^2}\, \left\langle {\cal O}_8^{H}(^3P_0) \right\rangle 
\ ,
\label{MaHdef}
\end{equation}
where $a \approx 3.5$. 
%
\begin{table}
\begin{center}
\begin{tabular}{|c|c|c|c|c|}
\cline{2-5}
\multicolumn{1}{c|}{}
&  $ \left\langle {\cal O}_1^{H}(^3S_1) \right\rangle $
&  $ \left\langle {\cal O}_1^{H}(^3P_1) \right\rangle /\mc^2 $
&  $ \left\langle {\cal O}_8^{H}(^3S_1) \right\rangle $
&  $ {\cal M}_{3.5}^{H} $
\\[1ex]\hline 
$\psi(nS)$ & $1$ & $v^6$ & $v^4$ & $v^4$\\ \hline
$\JP$ & 1.16  & - & $1.1\times 10^{-2}$ & $4.4 \times 10^{-2}$
\\ \hline
$\psi'$ & 0.76  & - & $3.8\times 10^{-3}$ & $1.8 \times 10^{-2}$
\\ \hline\hline
$\chi_{\c J}$ & $v^6$ & $v^2$ & $v^2$ & $v^4$\\ \hline
 $\chi_{\c 1}$ & -  & $0.32/1.48^2$ & $2.4\times 10^{-2}$ & -
\\ \hline
\end{tabular}
\end{center}
\caption[]{Scaling of NRQCD matrix elements with $v$ (modulo $v^3$) 
and numerical values in GeV$^{-3}$ as obtained from a potential-model
calculation (second and third row) \cite{EQ}
and fits to the Tevatron $\pT$ 
distribution (last two rows) \cite{ref:tevatron}.
\label{tab:MEone}}
\end{table}
The numerical values are indeed consistent with the expectation from
velocity scaling, see table~\ref{tab:MEone}.
The values are, however, subjected to a number of
uncertainties:
\begin{itemize}
\item
  The value of the charm-quark mass; it affects mainly the overall 
  normalization and should always be supplied when quoting
  values for the ME.
\item
  The parametrization of the parton-distribution functions (PDF) 
  and the choice 
  of the factorization scale; both influence the normalization as well as
  the shape of the $\pT$ distribution. 
\item
  The value of 
  $\LQCD$ (which is not independent of the PDF) 
  and the renormalization scale $\mu_R$; note that processes (\ref{ppone}) and
  (\ref{pptwo}) start at high order in $\as$. The values of
  $\LQCD$ and $\mu_R$ affect also the
  $\pT$ shape, since $\mu_R^2 \sim m^2 + \pT^2$. 
\item
  The lack of 
  higher-order perturbative QCD corrections. Particularly important are: 
  (i) initial-state radiation and intrinsic $\kT$ effects since they may 
  modify the slopes of (\ref{ppone}) and
  (\ref{pptwo}) considerably \cite{Cano}; 
  (ii) Colour-singlet production via the $O(\as^4)$ reaction 
  $\g + \g \rightarrow \c\cbar_1({}^3S_1) + \g + \g$, since it
  has the same $\pT$ fall-off 
  as (\ref{ppone}) \cite{Schuler}.
\item
  Higher-order $v^2$ corrections are important close to boundaries of 
  partonic thresholds \cite{BRW}.
\item
  Higher-twist corrections. At large $\pT$ these are probably small 
  for the short-distance cross sections. However, sizeable corrections
  may be present for the fragmentation functions at the input scale.
\item
  Values of the colour-singlet matrix elements; at the Tevatron this is
  important for $\chi_{\c J}$ production only, i.e.\ the extracted value of
  $\langle {\cal O}_8^{\chi_{\c J}}(^3S_1) \rangle$.
\end{itemize}
From the above discussion we expect the current determination 
(table~\ref{tab:MEone}) of 
$\langle {\cal O}_8^{\psi(nS)}(^3S_1) \rangle$ to be on the low side, 
$M_{3.5}^{\psi(nS)}$ to be overestimated, 
and a strong correlation between 
$\langle {\cal O}_8^{\chi_{\c 1}}(^3S_1) \rangle$ and
$\langle {\cal O}_1^{\chi_{\c 1}}(^3P_1) \rangle$.

Although it is reassuring that the FA can accommodate the data, 
the result is not too surprising given that we have
two additional mechanisms that scale as $1/\pT^4$ and $1/\pT^6$,
respectively, with free normalizations. Additional consistency checks
are therefore needed. The polarization of the $\JP$, measurable 
via its decay into $\mu^+\mu^-$, is one such test since it is 
uniquely predicted without the need of additional long-distance MEs 
up to $O(v^4)$ \cite{Cacciari}.
Polarization measurements should soon become available
at the Tevatron. Here we focus on another aspect of the FA, 
namely the universality of the ME. To this end we compare MEs extracted
from various reactions.

\section{Z-decay}
\begin{figure}
\begin{center}
\epsfig{file=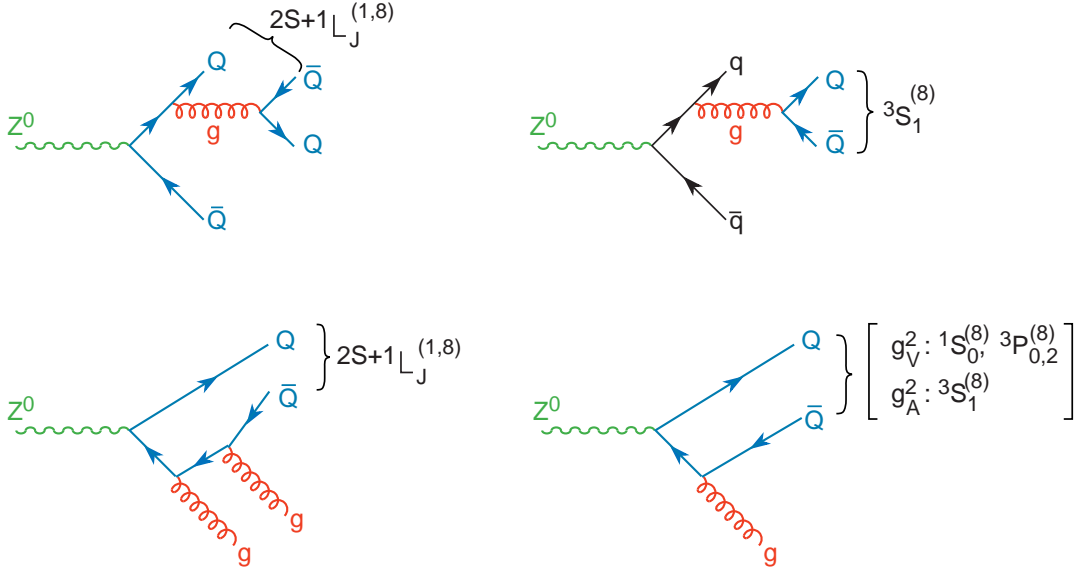,width=0.98\textwidth}
\end{center}
\caption{Feynman diagrams for quarkonium production in $\Z^0$ decays.
\label{fig:Z}}
\end{figure}
The Feynman diagrams for quarkonium production
in $\e^+\e^-$ annihilation are shown in Fig.~\ref{fig:Z}. 
Results of a recent prediction \cite{IJMPA} are compared with LEP data
\cite{LEP-charm,LEP-bottom} in table~\ref{tab:Zdecays}. 
At $\e^+\e^-$ c.m.\ energies $\sqrt{s}$ that are large with respect 
to the heavy-quark mass, 
for example in $\Z^0$ decays, 
contributions with additional gluons (lower diagrams in Fig.~\ref{fig:Z})
are suppressed by powers of $m/\sqrt{s}$.
The dominant 
contributions arise from $Q\bar{Q}$ production in the leading 
(colour-singlet) Fock state (upper left diagram in Fig.~\ref{fig:Z}) 
and the production of a $Q\bar{Q}_8(^3S_1)$ pair through gluon splitting 
(upper right diagram in Fig.~\ref{fig:Z}). The former contribution 
is the only one present in the so-called colour-singlet model (CSM). 
As table~\ref{tab:Zdecays} reveals, this colour-singlet contribution alone 
seems low with respect to the $\JP$ data, although still compatible,  
within errors. 

The data are well described, after inclusion of the colour-octet part, 
with a value for $\langle {\cal O}_8^{\JP}(^3S_1) \rangle$ as extracted 
from fits to the Tevatron $\pT$ distribution. 
In the case of $\Upsilon$ production, the statistics per experiment 
is too low to allow for definite conclusions. 
Further improvement can be expected because not all $\Z^0$ data 
have yet been analysed. Since the error of the data is 
statistics-dominated, a combined LEP analysis would be welcome. 
Information on colour-octet MEs other than 
$\langle {\cal O}_8^{\JP}(^3S_1) \rangle$ will come only from 
$\e^+\e^-$ annihilation at lower energies, BES, CESR, and $\B$-factories. 

The LEP data are also consistent with the expectation of the
colour-evaporation model (CEM), a model of quarkonium formation that 
assigns precisely one process-independent, long-distance factor $f^H = 
\langle {\cal O}^{H}[n] \rangle$ to each quarkonium particle $H$ \cite{GaS}.
The cross section $\sigma(Q\bar{Q}[n])$, see (\ref{factcross}), 
is obtained by integrating the $Q\bar{Q}$ production cross section 
up to the threshold of the production of a pair of heavy-light mesons. 
With parameters $f^H$ extracted from fixed-target hadroproduction, the CEM 
is also able to describe the Tevatron $\pT$ distributions \cite{Vogt}. 
\begin{table}
\begin{center}
\begin{tabular}{|c|ccc|cccc|}
\cline{2-8}
\multicolumn{1}{c|}{}
 & CSM & CEM & NRQCD & DELPHI & OPAL & ALEPH & L3
\\ \hline
$\JP$ & $7.8$ & $23$ & $26$ & $44^{+36}_{-30}$ & $19\pm 10$ & $30\pm 9$
 & $27\pm 12$
\\ \hline
$\sum \Upsilon$ & $1.7$ & $1.7$ & $6.6$ 
& $< 124$ & $10\pm 5$ & $< 7.3$ & $< 7.6$
\\ \hline
\end{tabular}
\end{center}
\caption{Branching ratios and $95\%\,$CL upper limits
 in $10^{-5}$ of $\Z^0$ decays: production of the 
sum of the three lowest-lying $\Upsilon$ states 
and prompt $\JP$ production (i.e.\ excluding
$\B$ decays but including feed-down from $\chi_{\c J}$ and $\psi'$).
\label{tab:Zdecays}}
\end{table}

\section{Low-energy electron--positron annihilation}
\begin{figure}[tbp]
\begin{center}
\epsfig{file=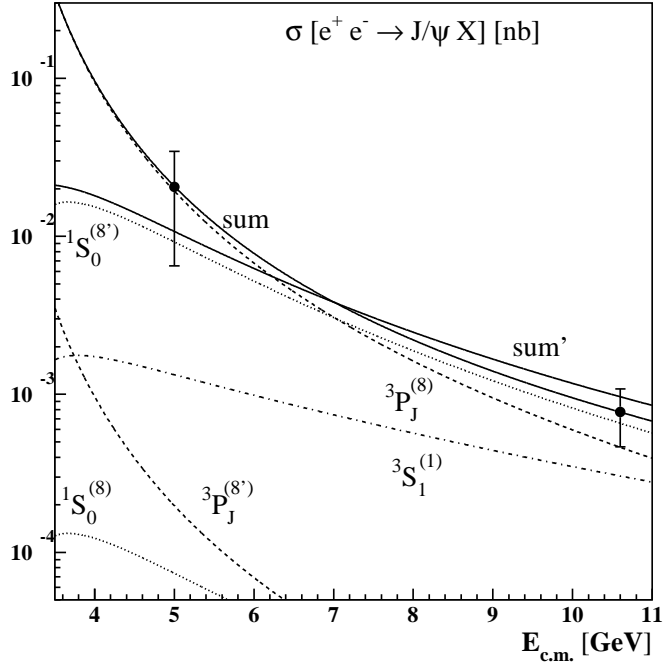,width=0.60\textwidth}
\end{center}
\caption{Fits to direct $\JP$ production in $\e^+\e^-$ annihilation.
\label{fig:lowee}}
\end{figure}

In low-energy $\e^+\e^-$ annihilation, $\JP$ production is unique
in the sense that it is the only process known so far where
the colour-octet ${}^1S_0$ and ${}^3P_0$ MEs enter with widely
different weights. At low energies, the dominant contributions involve 
a $\c\cbar$ pair plus one or two perturbative gluons 
(lower Feynman diagrams in Fig.~\ref{fig:Z}) \cite{eeref,YQC}.
\begin{table}
\begin{center}
\begin{tabular}{|c|c|c|c|c|}
\hline
$\sqrt{s}$ & $R$[exp.] & $R_1$  & $R_8(^1S_0)$ & $R_8(^3P_0)$
\\ \hline
$5\,{\mathrm GeV}$ & $60\pm 40$ & 
  $3.3\, \langle {\cal O}_1^{\JP}(^3S_1)\rangle$ &
  $530\, \langle {\cal O}_8^{\JP}(^1S_0)\rangle$ &
  $\frac{6000}{\mc^2}\, \langle {\cal O}_8^{\JP}(^3P_0)\rangle$ 
\\ \cline{3-5}
 & & 
  $3.8$ & $0.2$ & $56$
\\ \cline{2-5}
 & $20$ & $3.8$ & $39$ & $-23$
\\ \hline \hline
$10.6\,{\mathrm GeV}$ & $10\pm 4$ & 
  $3.4\, \langle {\cal O}_1^{\JP}(^3S_1)\rangle$ &
  $170\, \langle {\cal O}_8^{\JP}(^1S_0)\rangle$ &
  $\frac{640}{\mc^2}\, \langle {\cal O}_8^{\JP}(^3P_0)\rangle$ 
\\ \cline{3-5}
 & &  $3.9$ & $0.07$ & $6.0$
\\ \cline{2-5}
 & $14$ & $3.9$ & $12.5$ & $-2.5$
\\ \hline
\end{tabular}
\end{center}
\caption{Cross section of prompt $\JP$ production in $\e^+\e^-$ 
annihilation (excluding $\b$ decays) in units of $10^4 \sigma_{\mu\mu}$:
Data \cite{Pluto} ($R$[exp.]), 
colour-singlet contribution ($R_1$), 
and colour-octet contributions through ${}^1S_0$ 
and ${}^3P_J$ intermediate $\c\cbar$ pairs. 
The PLUTO data $\sigma = (31\pm 21)\,$pb
have been multiplied by $(1-0.23)\, 6/7$ to remove the $\psi'$ 
contribution and to update the leptonic branching ratio $\JP \rightarrow
\mu^+ \mu^-$. 
The theoretical values are for $\as = 0.278$, $\mc=1.5\,$GeV, 
the colour-singlet ME as given in table~\ref{tab:MEone} and 
the colour-octet MEs (\ref{octetee}).
Also given are the results of one of the extreme solutions (\ref{extreme}).
\label{tab:ee}}
\end{table}
The cross sections are readily evaluated and results presented 
in table~\ref{tab:ee} and Fig.~\ref{fig:lowee}.

Observe first that the ME combination 
(\ref{MaHdef}) enters with $a$ varying between $11.3$ and $3.8$ 
in $5 < \sqrt{s} < 10.6\,$GeV. The central values of the experimental data
yield for $\as=0.278$, $\mc = 1.5\,$GeV, and 
$\langle {\cal O}_1^{\JP}(^3S_1)\rangle = 1.16\,$GeV$^3$
\begin{equation}
  \left\langle {\cal O}_8^{\JP}(^1S_0) \right\rangle 
  = 4.0\times 10^{-4}\,{\mathrm GeV}^3\ , \qquad
 \frac{ \left\langle {\cal O}_8^{\JP}(^3P_0) \right\rangle }{\mc^2}
  = 9.3\times 10^{-3}\,{\mathrm GeV}^3\ , 
\label{octetee}
\end{equation}
and hence
${\cal M}_{3.5}^{\JP}  = 3.3 \times 10^{-2}\,{\mathrm GeV}^3$ 
(unprimed curves in Fig.~\ref{fig:lowee}).
The last number is somewhat smaller than the naive Tevatron estimate, 
see table~\ref{tab:MEone}. 

Our numbers are similar to those found in \cite{YQC}. There it was claimed
that $\langle {\cal O}_8^{\JP}(^3P_0) \rangle/\mc^2$ can be extracted very 
precisely from low-energy $\e^+\e^-$ data while 
$\langle {\cal O}_8^{\JP}(^1S_0) \rangle$ is hardly constrained, 
based on the dominance of $R_8(^3P_0)$ over $R_1$ and $R_8(^1S_0)$ 
at low energies and the fact that $R_8(^1S_0)$ is a small contribution
for all $\sqrt{s}$. Indeed, varying $\as$ in $(0.24,0.30)$, 
$\mc$ in $(1.4,1.6)\,$GeV, and $\langle {\cal O}_1^{\JP}(^3S_1) \rangle$
in $(0.80,1.52)\,$GeV$^3$ yields, for the central experimental values, 
the following ranges\footnote{%
These are the maximal ranges obtained by varying one parameter at a time; 
the uncertainty associated with each of the three parameters is about 
the same.}:
\begin{equation}
  -1.2 < \frac{10^2}{{\mathrm GeV}^3} \,
    \left\langle {\cal O}_8^{\JP}(^1S_0) \right\rangle  <  1.2
\ , \quad
   1.1 > \frac{10^2}{{\mathrm GeV}^3\, \mc^2} \,
    \left\langle {\cal O}_8^{\JP}(^3P_0) \right\rangle  > 0.78
\ ,
\label{theorextreme}
\end{equation}
implying 
$ 2.6 < 10^2\, {\cal M}_{3.5}^{\JP} / {\mathrm GeV}^3 < 4.5$. 

However, a few remarks are in order. First, the large experimental errors 
forbid definite conclusions\footnote{%
Recall that PLUTO \cite{Pluto} observes an excess of only $4$
prompt $\e^+\e^- \rightarrow \JP\, X$ events; out of these, only
the two events with both decay muons reconstructed
have been used here.}. 
Even for fixed theoretical parameters
$\as$, $\mc$, $\langle {\cal O}_1^{\JP}(^3S_1)\rangle$ 
only the following, much wider range is obtained:
\begin{equation}
  -0.73 < \frac{10^2}{{\mathrm GeV}^3} \,
    \left\langle {\cal O}_8^{\JP}(^1S_0) \right\rangle  <  7.4
\ , \quad
   2.2 > \frac{10^2}{{\mathrm GeV}^3\, \mc^2} \,
    \left\langle {\cal O}_8^{\JP}(^3P_0) \right\rangle  > -0.38
\ ,
\label{extreme}
\end{equation}
implying 
\begin{equation}
    {\cal M}_{3.5}^{\JP} = \left( 3.3 \pm 2.7 \right)\times 10^{-2}
    \,  {\mathrm GeV}^3\ .
\end{equation}
Insisting in $\langle {\cal O}_8^{\JP(^1S_0)} \rangle > 0$, the 
minimally allowed value of ${\cal M}_{3.5}^{\JP}$ increases from 
$0.57\times 10^{-2}\,$GeV$^3$ to $1.1\times 10^{-2}\,$GeV$^3$. 
One pair of MEs fulfilling (\ref{extreme})
($\langle {\cal O}_8^{\JP(^1S_0)} \rangle = 0.05\,$GeV$^3$,
$\langle {\cal O}_8^{\JP}(^3P_0) \rangle/\mc^2 = 9.5/10^5\,$GeV$^3$) 
is shown in Fig.~\ref{fig:lowee} (primed curves). 
If we allow in addition the (independent) variation of the theoretical 
parameters, even negative values of ${\cal M}_{3.5}^{\JP}$ are not excluded.

Negative values for $\langle {\cal O}_8^{\JP}(^3P_0) \rangle$ are 
not {\it per se} excluded, 
although at first sight these would lead to negative 
cross sections at low c.m.\ energies since the cross section of 
$\c\cbar_8(^3P_J)$ production grows at low energies while all others vanish. 
However, the $\e^+\e^- \rightarrow \JP X$ cross section at low energies 
is no longer given by the fixed-order calculation. 
The reason is the singular behaviour of 
the ${}^3P_J$ cross sections: for $\xi = 2 \mc/\sqrt{s}  \rightarrow 1$,  
these diverge as $1/(1-\xi)$. Clearly this indicates the breakdown of 
the fixed-order in $v^2$ calculation: $v^4/(1-\xi)$ is no longer a small
parameter and higher-order terms in $v^2$ grow more rapidly as 
$\xi \rightarrow 1$. The energy dependence at low energies can only 
be described once these terms of higher order in $v^2$ are resummed. 
Eventually, as $1-\xi < v^2$, higher-twist contributions become 
important, too.
Of course, the result of the fixed-order calculation can still be used
if we smear over a sufficiently wide region in $\sqrt{s}$. 

Near threshold the cross section is given by
\begin{eqnarray}
\lefteqn{
  \sigma\left[ \e^+\e^- \rightarrow \JP + x\right]_{\mathrm direct} 
 \stackrel{\xi \rightarrow 1}{\longrightarrow} 
} \nonumber\\ & &
     \sigma_0\,     \left\{
     \left\langle {\cal O}_1^{\JP}(^3S_1) \right\rangle \, \left[ 
    \delta(1-\xi) + \left(\frac{\as}{\pi}\right)^2 \, \frac{32}{81}\,
      (1-\xi) \right]  
  \right. 
\nonumber\\ & &~ \left. + \frac{\as}{\pi}\, \left[ 8\, (1-\xi)\, 
     \left\langle {\cal O}_8^{\JP}(^1S_0) \right\rangle 
     + \frac{8}{3}\, \frac{1}{1-\xi}\, \frac{1}{\mc^2}\, 
    \sum_{J=0}^{2}\, \left\langle {\cal O}_8^{\JP}(^3P_J) \right\rangle 
    \right]  \right\}
\ ,
\end{eqnarray}
where $\sigma_0 = \pi^3 \alpha_{\mathrm em}^2 e_c^2/(6 \mc^5)$, 
$e_c = 2/3$.
The apparent singularity at $\xi=1$ can be absorbed through the 
factorization-scale dependence of the leading colour-singlet ME owing to the
evolution equation derived in \cite{BBL95,Mannel}
\begin{equation}
\Lambda\, \frac{\d }{\d \Lambda}\, {\cal O}_1(^3S_1) = 
\frac{8 \as}{3\pi\mc^2}\, \sum_{J=0}^2\, {\cal O}_8(^3P_J)
  - \frac{8C_F\as}{3\pi\mc^2}\, {\cal P}_1(^3S_1)
\ .
\end{equation}
Hence a finite cross section is arrived at when averaging 
the threshold region over a range of order $\mc$.

\section{B decay}
The $\B$ meson is known to decay into charmonia with branching ratios 
between $0.1$ and $1\%$ \cite{CLEO}. 
Branching ratios into charmonia of the other 
$\b$-flavoured hadrons are not known. 
The hard process is given by
\begin{equation}
 \b \rightarrow \c\cbar_c[n] + X_s
\, , \qquad n = {}^1S_0\, , \; {}^3S_1\, , \; {}^3P_1\, , \; 
 D{\mathrm -waves}\, , \ldots
\ ,
\label{b-decay}
\end{equation}
where $X_s$ is a system containing a strange quark and, as usual, 
the subscript $c$ indicates a colour-singlet ($c=1$) 
or colour-octet ($c=8$) heavy-quark pair. 
Calculations of $\B$ decays into charmonium states 
based on (\ref{b-decay}) are valid to leading order in $v^2$
and leading order in $\LQCD/\mb$. Using the results of
\cite{Brefs}, we can find the $\B$ into charmonium decay-branching ratios
in terms of the semileptonic branching ratio
\begin{eqnarray}
  {\mathrm Br}\,[\B \rightarrow H + X] & = & 
  {\mathrm Br}\,[\b \rightarrow \c\, \ell\, \nu_{\ell}] \, 
  \frac{4\, \pi^2}{9\, \mb^3}\, 
  \frac{R(\mc/\mb)}{f(\mc/\mb)}
\nonumber\\ & &~ \times
  \left\{ \left[ 2\, C_+ - C_- \right]^2\, \hat{\Gamma}_1^H
     + \left[ C_+ + C_- \right]^2\,
     \frac{3}{2}\, \hat{\Gamma}_8^H \right\}
\ .
\label{BBr}
\end{eqnarray}
Here $f(r)$ is the  phase-space function 
for the $\b$ semileptonic decay
\begin{equation}
f(r) = 1-8\,{r}^{2}+8\,{r}^{6}-{r}^{8}-24\,{r}^{4}\ln (r)
\ ,
\end{equation}
$R(z)$ is the phase-space function for $\b \rightarrow H + X$,
\begin{equation}
  R(z) = \frac{1}{z}\, \left[ 1 - (2\, z)^2 \right]^2 
\ ,
\label{Rdef}
\end{equation}
and
\begin{eqnarray}
  \hat{\Gamma}_c^H & = & 
  (1 + 8\, z^2)\,   \left\langle {\cal O}_c^{H}(^3S_1) \right\rangle 
  - \frac{1}{3\, \mc^2}\, \left\langle {\cal P}_c^{H}(^3S_1) \right\rangle 
\nonumber\\ & &~
  + 3\,   \left\langle {\cal O}_c^{H}(^1S_0) \right\rangle 
  +  (1 + 8\, z^2)\,  \frac{2}{\mc^2}\,  
     \left\langle {\cal O}_c^{H}(^3P_1) \right\rangle 
\ .
\label{hatGamma}
\end{eqnarray}
Of course, 
for a particular charmonium particle $H$ and a specified colour state $c$ 
not all MEs in (\ref{hatGamma}) contribute to any given order in $v^2$, 
see table~\ref{tab:MEone}. The second ME in (\ref{hatGamma}),
$\langle {\cal P}_c^{H}(^3S_1) \rangle$, is a $v^2$ correction 
to $\langle {\cal O}_c^{H}(^3S_1) \rangle$ and will be 
neglected in the following.

Finally, $C_{\pm}$ are factorization-scale-dependent Wilson coefficients 
describing the evolution from the W-mass scale down to a scale of the
order of the $\b$-quark mass.
Note that the coefficient of the colour-octet contribution is strongly 
enhanced compared with the colour-singlet one:
\begin{equation}
  \frac{3}{2}\, 
  \left( \frac{ C_+ + C_- }{2\, C_+ - C_- } \right)^2 \approx 50
\ .
\label{Wilsonratio}
\end{equation}
In the following we take as central values 
$C_+ + C_- = 2.2$, $2\, C_+ - C_- = 0.38$, 
$\mc = 1.48\,$GeV, $\mb = 5.0\,$GeV, 
${\mathrm Br}\,[\b \rightarrow \c\, \ell\, \nu_{\ell}] = 10.3\,$\%. 
In contrast to the colour-octet Wilson coefficient, 
the colour-singlet one is strongly scale sensitive, since it is given as
the difference of two large numbers. In order to bracket its uncertainty
we shall vary $2\, C_+ - C_-$ between $0.19 = 0.38/2$ and 
$0.645$. With the latter value, Br$\,[\B \rightarrow \JP X]$ is
saturated with only the colour-singlet contribution (see below). 
Varying the colour-singlet coefficient in $(0.19,0.645)$ gives the range 
from $17$ to $201$ for the ratio (\ref{Wilsonratio}).

As is true for most quarkonium production reactions, 
$\B$ decays into charmonia are sensitive to only the combination 
(\ref{MaHdef}) of colour-octet ME. 
Moreover, the value of $a$ is similar to the one at the Tevatron, 
$a = 2\, (1 + 8\, z^2) \approx 3.4$. 
This follows immediately upon using the approximate symmetry relation
$  \langle {\cal O}_8^{H}(^3P_1) \rangle \approx 3\, 
  \langle {\cal O}_8^{H}(^3P_0) \rangle $.

Restricting to the lowest order in $v^2$ terms, separately for the
colour-singlet and colour-octet contributions, we find the following
{\em direct} branching ratios
\begin{equation}
\begin{array}{cccl}
 {\mathrm Br}\,[ \B \rightarrow \psi + X] = & 
  0.24\, \frac{
  \left\langle {\cal O}_1^{\psi}(^3S_1) \right\rangle }{ {\mathrm GeV}^3 }
 &  +~ 12 \, \frac{ 
  \left\langle {\cal O}_8^{\psi}(^3S_1) \right\rangle }{ {\mathrm GeV}^3 } &
  +~
  21 \, \frac{
  {\cal M}_{3.4}^{\psi} }{ {\mathrm GeV}^3 }
\\
 \JP: 0.80 \pm 0.08 & 0.28\ (0.07,0.80)   & 0.13~ & 0.92
\\
 \psi': 0.34 \pm 0.05 & 0.18\ (0.04,0.52) & 0.046 & 0.38 
\\
                      & 0.10 & 0.024 & 0.20 \quad (\kappa=0.53)
\end{array}
\end{equation}
and
\begin{equation}
\begin{array}{cccl}
 {\mathrm Br}\,[ \B \rightarrow \chi_{\c J} + X] = &
  0.48\, \frac{
  \left\langle {\cal O}_1^{\chi_{\c 1}}(^3P_1) \right\rangle }%
    { \mc^2 {\mathrm GeV}^3 } \, \delta_{1,J} 
 &  +~ 12 \, \frac{ 
  \left\langle {\cal O}_8^{\chi_{\c J}}(^3S_1) \right\rangle }%
    { {\mathrm GeV}^3 } &
\\
 \chi_{\c 1}: 0.37 \pm 0.07 
  & 0.070\ (0.017,0.20) & 0.28 & 
\\
                            & 0.046 & 0.19 &  \quad (\kappa=0.66)
\\
 \chi_{\c 2}: 0.23 \pm 0.10 &     0 & 0.47 & 
\\
                             &     0 & 0.29 &  \quad (\kappa=0.63)
\end{array}
\label{Bchi}
\end{equation}
The experimental numbers (first rows) are the CLEO measurements \cite{CLEO}
of the direct branching ratios (obtained by subtracting the feed-down  
from $\psi' \rightarrow \chi_{\c J}\, X$ and 
$\chi_{\c J}\rightarrow \JP\, \gamma$). 
The numbers in brackets denote the range of the colour-singlet contribution
obtained by varying $2C_+-C_-$ in the range specified above. 
The numbers quoted for $\kappa \neq 1$ are obtained by using meson masses 
rather than quark masses in the phase-space function, i.e.\ by multiplying 
(\ref{BBr}) by $\kappa = R([m_H/2] / m_{\B}) / R(\mc/\mb)$. 
Clearly there are other $v^2$, $\LQCD/\mb$, and even
$\LQCD/\mc$ corrections besides this trivial phase-space factor. 
Hence these numbers should be regarded as an only very naive indication 
of the uncertainties due to these corrections.

We observe that, by stretching the poorly determined 
colour-singlet Wilson coefficient $2C_+ - C_-$, production of 
$\JP$ and $\psi'$ can be explained merely by
the colour-singlet contribution, in contrast to $\chi_{\c J}$
production. This is in accord with the fact that there are two 
$\chi_{\c J}$ MEs in leading $v^2$ while only one $\psi$ ME survives
for $v \rightarrow 0$. On the other hand,  
the values of the $\psi$ ${}^3S_1$ colour-octet MEs are well compatible 
with the CLEO data for almost the whole range of 
$2C_+ - C_-$. In any case, 
the colour-octet ME combination ${\cal M}_{3.4}^{\psi}$ is too large, 
for example, by a factor of about $2$ 
for the central value of $2C_+ - C_-$. 
This confirms the expectation raised above that it has been overestimated
in fits to the Tevatron data. 

An inspection of the $\chi_{\c 2}$ width reveals that the
$\chi_{\c J}$ ${}^3S_1$ colour-octet ME is a factor of about $2$ too large.
The $\chi_{\c 1}$ width then requires the colour-singlet ME
$\langle {\cal O}_1^{\chi{\c 1}}(^3P_1) \rangle$ to be larger, 
by a factor of about $3$ for the central value of $2C_+ - C_-$. 
This may well be consistent with the Tevatron data where there is a
similar strong correlation between the two MEs. 

While the leading-order $\chi_{\c J}$ ME is rather well known, 
MEs of higher orders in $v^2$ have not yet been determined. 
However, relativistic corrections are likely to be important for 
$\chi_{\c J}$ production in fixed-target experiments and 
in $\e^+\e^-$ collisions \cite{Mikko}. 
Corrections to $O(v^2)$ arise merely from operators with two 
extra spatial derivatives, but the same quantum numbers as the leading ones. 
Contributions that scale as $v^4$ relative to (\ref{Bchi}) give 
the correction
\begin{eqnarray}
 \Delta\, {\mathrm Br}\,[ \B \rightarrow \chi_{\c J} + X]  & = & 
 \left\{
  0.24\, 
  \left\langle {\cal O}_1^{\chi_{\c J}}(^3S_1) \right\rangle 
 \right.
   +  0.42\,
  \left\langle {\cal O}_1^{\chi_{\c J}}(^1S_0) \right\rangle 
\\ \nonumber & &~ 
  + 21 \, {\cal M}_{3.4}^{\chi_{\c J}} 
  \left.  + \frac{c_D}{\mc^4}\,  
  \left\langle {\cal O}_8^{\chi_{\c J}}(^3D_{J'}) \right\rangle 
  \right\}\, {\mathrm GeV}^{-3}
\ ,
\label{Btwochi}
\end{eqnarray}
where $c_D$ is an as yet uncalculated coefficient. Nonetheless, 
it is clear that $\B$ decays into $\chi_{\c J}$ states 
provide an upper limit on ${\cal M}_{3.4}^{\chi_{\c J}}$. 
Applying the scaling $v^4 \sim 0.1$ to the two leading-order MEs, 
we find values for ${\cal M}_{3.4}^{\chi_{\c 1}}$
varying between $2\times10^{-3}$ and $1.5\times 10^{-2}$. 
$\B$-decay data certainly favour values at the lower end. 

Let us finally mention that the $\chi_{\c 2}$-to-$\chi_{\c 1}$ 
ratio is a serious problem for the CEM, which predicts the 
ratio $5/3$. 

\section{Upsilon decay}
Inclusive charmonium production in $\Upsilon$ decays is another place
to test the NRQCD factorization approach. 
In principle, colour-octet contributions in both the $\b\bbar$ 
decay and the $\c\cbar$ production should be taken into account. 
However, 
relativistic corrections from the $\b\bbar$ in a colour-octet state 
turn out to be small \cite{Cheung}, in accordance with the naive expectation
of a suppression by $(v_{\b}/v_{\c})^4 \sim 1/10$
relative to a contribution having the $\c\cbar$ in a colour-octet state. 

A full calculation of $\Upsilon$ decays into charmonia does not yet exist.
The colour-singlet contribution to direct $\JP$ production starts
at $O(\as^6)$ and consists of the tree diagrams 
$\b\bbar_1(^3S_1) \rightarrow \c\cbar_1(^3S_1) + 4\g$ 
and the loop diagrams 
$\b\bbar_1(^3S_1) \rightarrow \c\cbar_1(^3S_1) + 2\g$ 
containing box diagrams. Only an estimate of the $O(\as^5)$ colour-singlet 
contribution to indirect $\JP$ production through $\chi_{\c J}$ decays 
exists, yielding about $1/20$ of the measured branching ratio 
\cite{Cheung,Trottier}. 

Colour-octet contributions to direct $\JP$ production in $\Upsilon$ decays
start at $O(\as^4)$ and are hence enhanced by $1/\as^2$. 
Although calculations of the colour-singlet 
contributions are still lacking, a comparison with data can be used to set 
upper limits on certain colour-octet MEs. 
The largest colour-octet contribution found so far is
$\b\bbar_1(^3S_1) \rightarrow \g\g\g^\star$ 
followed by $\g^\star \rightarrow \c\cbar_8(^3S_1)$ \cite{Cheung}. 
Another potentially large contribution is the loop process
$\b\bbar_1(^3S_1) \rightarrow \c\cbar_8(^1S_0,^3P_J) + \g$ 
proceeding through virtual gluons. This process can be 
related\footnote{While finishing this paper we learnt that 
this process had just been taken up also in \cite{Napsuciale}.}
to the radiative decay 
$Q\bar{Q}_1(^3S_1) \rightarrow \gamma + q\bar{q}_1(^{2S+1}L_J)$ 
calculated some time ago \cite{Korner}. 
We  find 
\begin{eqnarray}
{\mathrm Br}\, [\Upsilon \rightarrow \JP + X] & = & {\mathrm BR}_1 + R\, 
 \frac{\pi\, \as}{8\, (\pi^2-9)\, \mc^3}\, 
\nonumber\\ & &~ \quad
\left\{ 0.571\, \left\langle {\cal O}_8^{\JP}(^3S_1) \right\rangle 
 + 0.90\, \frac{\as}{\pi}\, {\cal M}_{3.8}^{\JP} \right\}
\nonumber\\
  & = & {\mathrm BR}_1 + 1.76\times 10^{-2}\, 
       \left\langle {\cal O}_8^{\JP}(^3S_1) \right\rangle 
 + 0.247\times 10^{-2}\, {\cal M}_{3.8}^{\JP} 
\nonumber\\
\left(11 \pm 4\right)\times 10^{-4} & = & {\mathrm BR}_1 +
  \left\{ 1.94  + 1.09 \right\}\, \times 10^{-4}
\ ,
\end{eqnarray}
where ${\mathrm BR}_1$ denotes the missing part from 
unknown colour-singlet contributions and indirect $\JP$ production. 
The second equation has been obtained
with $R = {\mathrm Br}\, [\Upsilon \rightarrow 3\g] \approx 0.83$,
$\as = 0.278$, $\mc = 1.5$. The number on the LHS of the last line 
is the CLEO measurement \cite{CleoY} and the values of the RHS follow from
insertion of the ME values in table~\ref{tab:MEone}. 
The latter are consistent with the measurement if, as expected, 
the colour-singlet part is small and the indirect
contributions does not exceed, say, $50\%$ of the total $\JP$ yield. 
More definite conclusions can only be drawn once the missing pieces 
have been calculated and/or more experimental information is available, 
for example, on the $\JP$ energy distribution or on the 
$\chi_{\c J}$ production rate. 

\section{Photoproduction}
\begin{figure}
\begin{center}
\epsfig{file=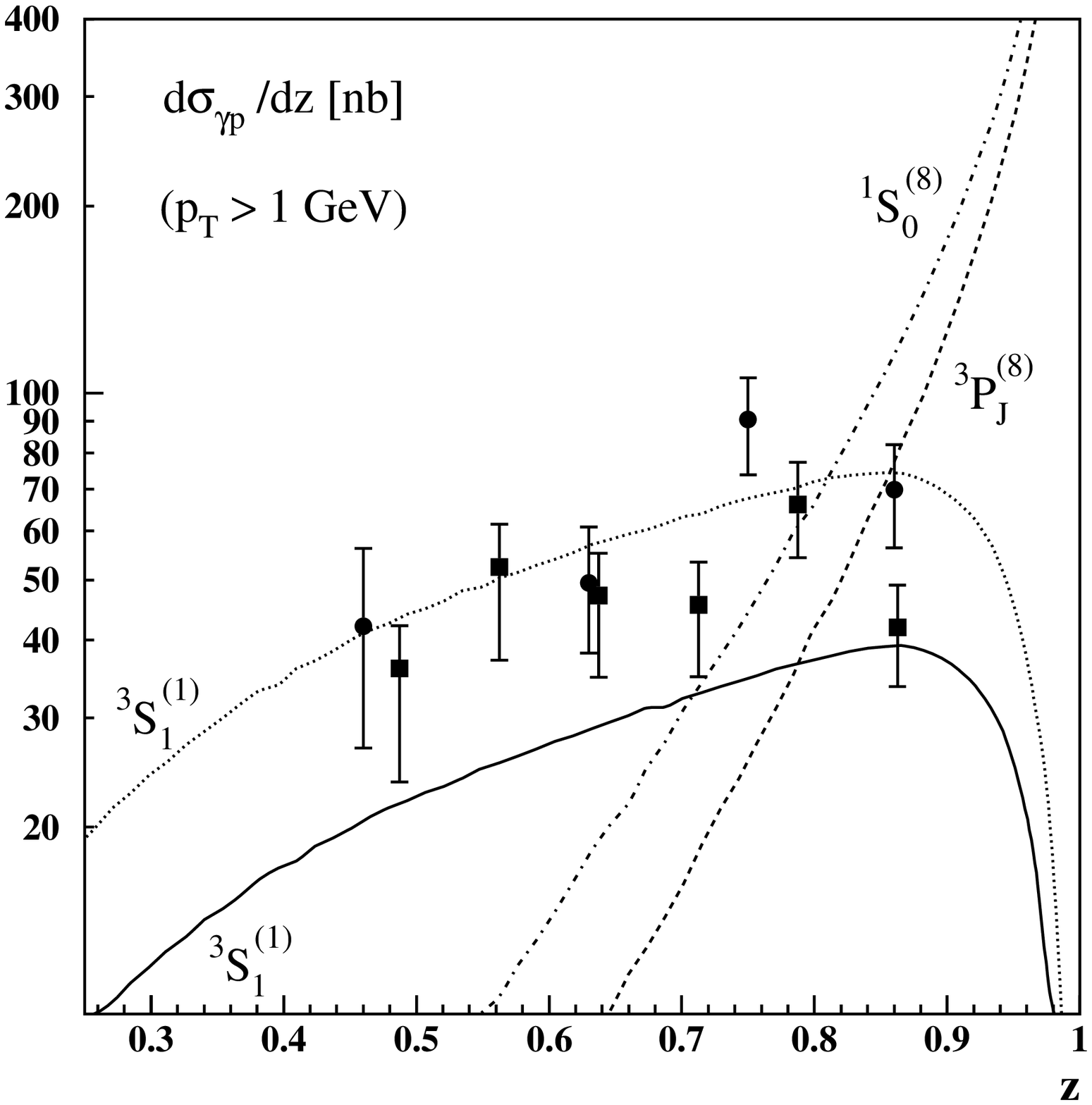,width=0.60\textwidth}
\end{center}
\caption{Photoproduction of $\JP$ compared with HERA\cite{HERA} data.
Theoretical curves at $\sqrt{s}_{\gamma \p}=100\,$GeV for
$\langle {\cal O}_1^{\JP(^3S_1)} \rangle = 1.16\,$GeV$^3$, 
$\langle {\cal O}_8^{\JP(^1S_0)} \rangle = 0.03\,$GeV$^3$,
$\langle {\cal O}_8^{\JP}(^3P_0) \rangle = 9/10^3\,$GeV$^3$, 
$\as = 0.278$, and $\mc=1.5\,$GeV, using 
the leading-order GRV\cite{GRV} parton distributions with $\mu=2 \mc$. 
The colour-singlet contribution is shown also for 
$\as = 0.3$, $\mc=1.4\,$GeV (dotted line).
\label{fig:Hera}}
\end{figure}

Photoproduction of $\JP$ has been claimed to be in conflict 
with NRQCD factorization \cite{ref:zdist}. In fact, the 
short-distance production of merely the leading colour-singlet state via
\begin{equation}
  \gamma + \g \rightarrow \c\cbar_1(^3S_1) + \g
\label{gp:sing}
\end{equation}
yields a shape of the differential cross section $\d \sigma / \d z$,  
in good agreement with the HERA data \cite{HERA} (Fig.~\ref{fig:Hera}), 
and also in agreement with low-energy data \cite{Jung}. 
Here $z$ is the fraction of the photon energy carried by the $\JP$ in 
the proton rest frame. 

The absolute normalization suffers from uncertainties similar 
to the ones listed in section~\ref{sec:tevatron}
for $\JP$ production at the Tevatron:
value\footnote{Note that the $\mc$ dependence is partly
compensated by corresponding changes in the ME; this has not been taken
into account in Fig.~\ref{fig:Hera}.}
of $\mc$, parametrization of the parton-distribution functions,
value of $\LQCD$, choice of factorization and renormalization 
scales\footnote{The scale dependence has been reduced through the 
inclusion of next-to-leading order perturbative corrections \cite{Zunft}.}, 
and initial-state transverse-momentum effects, i.e.\ effects arising
from the intrinsic transverse momentum $k_{\perp}$ of the gluon within 
the proton and initial-state gluon radiation. 
The latter effects mainly influence the
$\JP$ $\pT$-distribution at low $\pT$, but do affect
the $z$ distribution as well, mostly its normalizations, less so its shape.
This occurs in particular if 
a cut on $\pT$ is applied as low as $1\,$GeV, not much larger than 
$\langle k_{\perp} \rangle \sim 500\,$MeV. Recall that fixed-order 
calculations are reliable only at large $\pT$ or for $\pT$-integrated 
cross sections but do not adequately describe the shape at low $\pT$. 

Relativistic corrections of order $v^4$ arise from 
\begin{eqnarray}
  \gamma + \g & \rightarrow & \c\cbar_8[n] + \g \qquad 
      n = {}^1S_0,\,  {}^3S_1,\, {}^3P_{J}
\nonumber\\
  \gamma + \q & \rightarrow & \c\cbar_8[n] + \q \qquad 
      n = {}^1S_0,\,  {}^3S_1,\, {}^3P_{J}
\ ,
\label{gp:octet}
\end{eqnarray}
and from reactions involving the partonic content of the photon. 
Although such ``resolved-photon processes'' become more important as 
the c.m.\ energy increases, they still affect the small-$z$ region, only, 
say, below $0.3$ at HERA energies. The quark-initiated reactions 
in (\ref{gp:octet}) are small with respect to gluon-initiated ones
at HERA energies.  Moreover, 
$\gamma \g \rightarrow \c\cbar_8(^3S_1) + \g$ 
is identical to (\ref{gp:sing}), but down in magnitude by a factor of 
about $50$.
Most important are therefore the other two gluon-initiated reactions in 
(\ref{gp:octet}). In fact, their contributions seem to be in clear 
conflict with the HERA data when using the 
MEs as given in table~\ref{tab:MEone}, see Fig.~\ref{fig:Hera}. 
This is not an artefact of the particular values of the $S$- and
$P$-wave MEs chosen in Fig.~\ref{fig:Hera}, since both 
contributions are rather similar in shape. 
(To be precise, the value of $a$ in the ME combination (\ref{MaHdef}) 
is $a = 4.7 \pm 1.4$ in the range $0<z<1$ for $\pT > 1\,$GeV.)

Three reasons make us believe that we do, in fact, 
not encounter a breakdown of 
NRQCD factorization, namely (i) the size of the ME, the treatment
of (ii) the hard process and (iii) the hadronization.
First, we have seen that all but the Tevatron estimates 
prefer smaller values of the ME combination ${\cal M}_{a\sim 5}^{\JP}$. 
And we have given arguments why 
the value needed to explain the Tevatron data is likely to be smaller.

Second, (\ref{gp:octet}) contains contributions that correspond to 
the evolution of the gluon (or quark) distribution functions. 
These terms arise from diagrams with $t$-channel gluon exchange and 
have to be removed via mass-factorization in calculations of fully
$\pT$-integrated cross sections \cite{NLO}. 
Cross sections integrated above some
$\pT$-cut are reliably described by the fixed-order (in $\as$) results 
based on (\ref{gp:octet}) only if the cut is sufficiently large. 
If the $\pT$ cut is as low as, say, $1\,$GeV then the average $\pT$ of 
the fixed-order calculation is considerably smaller than what would be
obtained from a proper calculation including soft-gluon (initial-state)
radiation. Since contributions from small $\pT$ lead, on average, 
to harder $z$ distributions and vice versa, we expect a flatter 
$z$ distribution after inclusion of initial-state radiation 
and intrinsic $k_{\perp}$. 

The third aspect concerns the sensitivity of the $z$ distribution 
to energy ranges in the $\c\cbar_8[n] \rightarrow \JP$ transition 
of the order of the binding energy (and smaller). The physical reason
is clear: the momentum carried away by light hadrons in the hadronization 
process makes it very unlikely that $z$ values close to $1$ can be reached. 
Qualitatively, the hadronization causes a smearing of the $\JP$ 
momentum over a region $\delta z \approx v^2 \approx 0.25$--$0.30$,
$\delta \pT \approx m v^2 \approx 0.5\,$GeV. 
Technically, the NRQCD expansion parameter at $z$ close to one 
is\footnote{Here
$\hat{z}$ is defined analogously to $z$, i.e.\ $\hat{z}$ ($z$) is 
the photon-energy fraction carried by the $\c\cbar$ pair (the $\JP$)
in the proton rest frame.}
$v^2/(1-\hat{z})$ rather than $v^2$ \cite{BRW}.
The amount of smearing caused by this breakdown of the fixed-order (in $v^2$)
calculation makes the latter unsuitable to predict the $z$ distribution. 
Predictivity can, however, be restored upon introduction of 
universal shape functions. 
Attempts computing these shape functions
in the Coulomb limit ($m\rightarrow \infty$) have started \cite{BSW}. 
If supplemented 
by a sensible continuation into the higher-twist region, a $z$ distribution 
is obtained in agreement with the HERA data, provided the 
above-mentioned initial-state problem is dealt with as well. 
The required colour-octet MEs need not be much smaller than the 
naive estimates in table~\ref{tab:MEone} \cite{BSW}.

\section{Conclusions}
The NRQCD factorization approach gives quarkonium production cross sections
as a (finite) sum of short-distance coefficients times long-distance MEs. 
For the factorization to hold, the latter have to be process-independent. 
In this paper we have tested the universality of charmonium MEs 
by comparing their values as extracted from various reactions. Particular
emphasis has been put to elucidate the uncertainties in such 
determinations. The results can be summarized as follows.

Up to now, the cleanest evidence for the need of short-distance
production of colour-octet states in the formation of $1^{--}$ mesons comes
from the $\JP$ and $\psi'$ $\pT$ distribution at the Tevatron.  
The numerical values of the corresponding colour-octet MEs are in line
with the expectation from velocity scaling. Our investigation 
suggests that 
$\langle {\cal O}_8^{\psi}(^3S_1) \rangle$ 
is somewhat larger and 
${\cal M}_{3.5}^{\psi}$ 
considerably smaller than currently extracted values 
(table~\ref{tab:MEone}; $\psi$ denotes $\JP$ or $\psi'$). 

In the case of $J^{++}$ mesons: a colour-octet ME, 
$\langle {\cal O}_8^{\chi_{\c 0}}(^3S_1) \rangle$,  
enters already at leading order in $v^2$. 
Its numerical value as extracted from the Tevatron is highly 
correlated with the value of the colour-singlet ME
$\langle {\cal O}_1^{\chi_{\c 1}}(^3P_1) \rangle$, whose value
is less well known from potentail-model calculations 
than the colour-singlet $\psi$ ME. Indeed, 
$\B$-meson decays into $\chi_{\c J}$ suggest a factor-of-$2$ smaller
$\langle {\cal O}_8^{\chi_{\c 0}}(^3S_1) \rangle$ value and a factor
of $3$ larger value for 
$\langle {\cal O}_1^{\chi_{\c 1}}(^3P_1) \rangle$.

Although the full calculation of $\Upsilon$ decays into charmonia 
does not yet exist, the decay $\Upsilon \rightarrow \JP + X$ 
does already provide upper limits on the $\JP$ colour-octet ME. 
If the colour-singlet contributions were really as small as suggested
by their $\as$ scaling, this would then not only prove that 
short-distance colour-octet production is at work but also result in
a $\langle {\cal O}_8^{\JP}(^3S_1) \rangle$ value larger than 
the current Tevatron estimate given in table~\ref{tab:MEone}. 

Similarly, $Z^0$ data at LEP are about a factor of $3$ above the 
prediction based on colour-singlet production alone. Inclusion 
of colour-octet processes with 
$\langle {\cal O}_8^{\JP}(^3S_1) \rangle$ as currently deduced 
from the Tevatron reconciles nicely the agreement with the LEP data. 

Confronting theory and experiment of electron--positron annihilation 
into $\JP X$ at two different low energies is in principle the best
way to separate the two colour-octet MEs
$\langle {\cal O}_8^{\JP}(^1S_0) \rangle$  
and
$\langle {\cal O}_8^{\JP}(^3P_0) \rangle$  
that enter, in practically all other reactions, only in the 
linear combination (\ref{MaHdef}) with almost constant $a$. 
However, current data suffer from statistics that are so low that 
they are compatible with pure colour-singlet production 
within 2 standard deviations. 
Neglecting the experimental errors, 
a value of ${\cal M}_{3.5}^{\JP}$ 
is found somewhat smaller than the Tevatron one (table~\ref{tab:MEone}). 

More accurate determinations of NRQCD MEs from $\e^+\e^-$ annihilation 
and $\Z^0$ decays are mainly hindered by low statistics, and improvements
can be expected with more data in the future. In contrast, 
constraints from $\Upsilon$ decays are currently limited by theory. 
The situation is similar for $\B$ decays into $\JP$ and $\psi'$: The largest
uncertainty arises from the Wilson coefficient
$2C_+ - C_-$ entering the effective weak Hamiltonian,
which even in next-to-leading order is theoretically known only poorly. 
Pushing it to its maximum value, $\psi$ production is compatible 
with pure short-distance colour-singlet production. For most of 
the uncertainty range, however, the current Tevatron estimates for
$\langle {\cal O}_8^{\psi}(^3S_1) \rangle$ are well compatible 
with data. On the other hand, ${\cal M}_{3.5}^{\psi}$ is once again 
found to be smaller than the Tevatron value in table~\ref{tab:MEone}, 
by about a factor of $2$.

Finally, photoproduction of $\JP$ as measured at HERA does not pose a 
problem to NRQCD factorization, once colour-octet MEs compatible
with all current data are being used, the $\pT$ broadening due to 
initial-state gluon radiation and intrinsic $k_{\perp}$ is taken 
into account, and the leading-order $v^2$ calculation is improved 
with the help of shape functions in order to extend its validity 
beyond a value of $z$ of about $0.7$. 

\noindent\\[2ex]
{\it Acknowledgement}\hfill\\
It is my pleasure to thank G.\ Buchalla and M.\ Beneke for fruitful
discussions.
This work was supported in part by the EU Fourth Framework Programme
``Training and Mobility of Researchers'', Network 
``Quantum Chromodynamics and the Deep Structure of Elementary Particles'', 
contract FMRX-CT98-0194 (DG 12- MIHT).


\begin{thebibliography}{99}
 
\bibitem{BBL95}
G.T.\ Bodwin, E.\ Braaten and G.P.\ Lepage, 
 \PRD{51} (1995) 1125.

\bibitem{BY93}
E.\ Braaten and T.C.\ Yuan, 
 \PRL{71} (1993) 1673.

\bibitem{pprefs}
M.\ Cacciari and M.\ Greco, \PRL{73} (1994) 1586;\hfill\\
E.\ Braaten, M.\ Doncheski, S.\ Fleming and M.\ Mangano, 
\PLB{333} (1994) 548;\hfill\\
D.P.\ Roy and K.\ Sridhar, \PLB{339} (1994) 141;\hfill\\
E.\ Braaten and S.\ Fleming, \PRL{74} (1995) 3327;\hfill\\
M.\ Cacciari, M.\ Greco, M.\ Mangano and A.\ Petrelli, 
\PLB{356} (1995) 560;\hfill\\
P.\ Cho and A.K.\ Leibovich, \PRD{53} (1996) 150, {\it ibid.\ } 6203;
\hfill\\
M.\ Beneke and M.\ Kr\"amer, \PRD{55} (1997) 5269.

\bibitem{Cacciari} For reviews see 
E.\ Braaten, S.\ Fleming and T.C.\ Yuan, {\it Ann.\ Rev.\ Nucl.\ Part.\
 Sci.\ } {\bf 46} (1996) 197;
\hfill\\
M.\ Beneke,  \hep{9703429}, 
 in {\it 24th Annual SLAC Summer Institute on Particle Physics: 
 The Strong Interaction, From Hadrons to Protons (SSI 96)}, 
 Stanford, USA,  1996, p.~549; 
\hfill\\
M. Cacciari,  \hep{9706374}, 
 talk given at {\it 32nd Rencontres de Moriond},
 Les Arcs, France, 1997.

\bibitem{ref:tevatron}
F.\ Abe et al., \PRL{79} (1997) 572, {\it ibid.\ } 578;
\hfill\\
A.\ Sansoni, for the CDF collab., \NPA{610} (1996) 373c. 

\bibitem{Cano}
B.\ Cano-Coloma and M.A. Sanchis--Lozano, 
\NPB{508} (1997) 753;
\hfill\\
P.\ Ernstr\"om, L.\ L\"onnblad and M.\ Vanttinen, \ZPC{76} (1997) 515;
\hfill\\
B.A.\ Kniehl and G.\ Kramer, \hep{9803256}.

\bibitem{Schuler}
G.A.\ Schuler, CERN-TH/7170-94, \hep{9403387}.

\bibitem{BRW}
M.\ Beneke, I.Z.\ Rothstein and M.B.\ Wise, \PLB{408} 
(1997) 373.

\bibitem{EQ} 
E.J.\ Eichten and C.\ Quigg, 
   \PRD{52} (1995) 1726.

\bibitem{IJMPA} G. A. Schuler, 
 {\it Int.\ J. Mod.\ Phys.\ } {\bf A12} (1997) 3951.

\bibitem{LEP-charm}
DELPHI collab., P.\ Abreu et al., \PLB{341} (1994) 109;
\hfill\\
M.\ Wadhwa for the L3 collab., talk given at {\it QCD97}, 
       Montpellier, 1997;
\hfill\\
ALEPH collab., contribution 624 to the {\it 1997 EPS-HEP conference}, 
Je\-ru\-sa\-lem, 1997; 
\hfill\\
OPAL collab., \PLB{384} (1996) 343. 

\bibitem{LEP-bottom}
DELPHI collab., P.\ Abreu et al., \ZPC{69} (1996) 575;
\hfill\\
ALEPH collab., contribution PA05-066 to the
 {\it EPS-HEP conference}, Warsaw, 1996;
\hfill\\
L3 collab., M.\ Acciarri et al., CERN-PPE/97-78, July 1997;
\hfill\\ 
OPAL collab., \PLB{370} (1996) 185.

\bibitem{GaS}
For a review see, for example, G.A.\ Schuler, \ZPC{71} (1996) 317.

\bibitem{Vogt}
G.A.\ Schuler and R.\ Vogt, \PLB{387} (1996) 181.

\bibitem{eeref}
 J.H.\ K\"uhn and H.\ Schneider, \ZPC{11} (1981) 263;\hfill\\
 W.-Y.\ Keung, \PRD{23} (1981) 2072;\hfill\\
 L.\ Clavelli, \PRD{26} (1982) 1610;\hfill\\
V.\ Barger, K.\ Cheung and W.-Y.\ Keung, \PRD{41} (1990) 1541;\hfill\\
 E.\ Braaten, K.\ Cheung and T.C.\ Yuan, \PRD{48} (1993) 4230;\hfill\\
V.M. Driesen, J.H. K\"uhn and E. Mirkes, \PRD{49} (1994) 3197;\hfill\\
P.\ Cho and A.K.\ Leibovich, 
  \PRD{54} (1996) 6690;\hfill\\
 P.\ Cho, \PLB{368} (1996) 171;\hfill\\
 K.\ Cheung, W.-Y.\ Keung and T.C.\ Yuan, \PRD{76} (1996) 877;\hfill\\
 S.\ Baek, P.\ Ko, J.\ Lee and H.S.\ Song, 
       \PLB{389} (1996) 609,  \PRD{55} (1997) 6839.

\bibitem{YQC}
F.\ Yuan, C.-F.\ Qiao, K.-T.\ Chao, \PRD{56} (1997) 321, {\it ibid.\ } 1663.

\bibitem{Pluto}
PLUTO collab., J.\ Burmester et al., \PLB{68} (1977) 283;
\hfill\\
R.A.\ Poling, for the CLEO collab., in  
{\it Proc.\ Joint Int.\ Lepton-Photon
Symposium \& Europhysics Conf.\ on High-Energy Physics}, Geneva, 
1991, eds.\ S.\ Hegarty et al.\ (World Scientific, Singapore, 1992), p.~546.

\bibitem{Mannel}
T.\ Mannel and G.A.\ Schuler, \ZPC{67} (1995) 159.

\bibitem{CLEO}
CLEO collab., R.\ Balest et al., \PRD{52} (1995) 2661.

\bibitem{Brefs}
P.\ Ko, J.\ Lee and H.S.\ Song, \PRD{53} (1996) 1409;
\hfill\\
S.\ Fleming et al., \PRD{55} (1997) 4098.

\bibitem{Mikko}
G.A.\ Schuler and M.\ V\"{a}nttinen, 
  CERN-TH-97-261, October 1997, \hep{9710227}. 

\bibitem{Cheung}
K.\ Cheung, W.-Y.\ Keung and T.C.\ Yuan, 
 \PRD{54} (1996) 929.

\bibitem{Trottier}
H.\ Trottier, 
 \PLB{320} (1994) 145.

\bibitem{Korner}
J.G.\ K\"orner, J.H.\ K\"uhn, M.\ Krammer and H.\ Schneider,
\NPB{299} (1983) 115.

\bibitem{Napsuciale}
M.\ Napsuciale, \hep{9710488}.

\bibitem{CleoY}
CLEO collab., R.\ Fulton et al., 
 \PLB{224} (1989) 445.

\bibitem{ref:zdist}
M.\ Cacciari and M.\ Kr\"amer, \PRL{76} (1996) 4128;
\hfill\\
P.\ Ko, J.\ Lee and H.S.\ Song, \PRD{54} (1996) 4312.

\bibitem{HERA}
H1 collab., S.\ Aid et al., \NPB{472} (1996) 3;
\hfill\\
ZEUS collab., J.\ Breitweg et al., \ZPC{76} (1997) 599.

\bibitem{Jung}
H.\ Jung, G.A.\ Schuler and J.\ Terron, 
  {\it Int.\ J.\ Mod.\ Phys.\ } {\bf A7} (1992) 7955. 

\bibitem{GRV}
M.\ Gl\"uck, E.\ Reya and A.\ Vogt, \ZPC{53} (1992) 127.

\bibitem{Zunft}
M.\ Kr\"amer, J.\ Zunft, J.\ Steegborn and P.M.\ Zerwas, 
  \PLB{348} (1995) 657; 
M.\ Kr\"amer, \NPB{459} (1996) 3.

\bibitem{NLO}
F.\ Maltoni, M.\ Mangano and A.\ Petrelli, 
\hep{9708349}; 
A.\ Petrelli, M.\ Cacciari, M.\ Greco, F.\ Maltoni and M.\ Mangano, 
 \NPB{514} (1998) 245.

\bibitem{BSW}
M.\ Beneke, G.A.\ Schuler and S.\ Wolf, CERN preprint in preparation.

\end{thebibliography}
\end{document}